# Ultrafast spin-to-charge conversion in antiferromagnetic (111)-oriented L1$_2$-Mn$_3$Ir

Huiling Mao,[1] Yuta Sasaki,[2] Yuta Kobayashi,[3] Shinji Isogami,[2]

Teruo Ono,[3,4] Takahiro Moriyama,[5,6] Yukiko K. Takahashi,[2] and Kihiro T. Yamada[1, a)]

[1)] *Department of Physics, Tokyo Institute of Technology, Tokyo 152-8551, Japan*

[2)] *National Institute for Materials Science, Tsukuba, Ibaraki 987-6543, Japan*

[3)] *Institute for Chemical Research, Kyoto University, Uji, Kyoto 611-0011, Japan*

[4)] *Center for Spintronics Research Network, Institute for Chemical Research, Kyoto University, Uji, Kyoto, 611-0011, Japan*

[5)] *Department of Materials Physics, Nagoya University, Furo-cho, Chikusa-ku, Nagoya 464-8603, Japan*

[6)] *PRESTO, Japan Science and Technology Agency, Kawaguchi, Saitama 322-0012, Japan*

[a)] *Author to whom correspondence should be addressed: yamada@phys.titech.ac.jp*

Antiferromagnetic L1$_2$-Mn$_3$Ir combines outstanding spin-transport properties with magnons in the terahertz (THz) frequency range. However, the THz radiation emitted by ultrafast spin-to-charge conversion via the inverse spin Hall effect remains unexplored. In this study, we measured the THz emission and transmission of a permalloy/(111)-oriented L1$_2$-Mn$_3$Ir multilayer by THz time-domain spectroscopy. The spin Hall angle was determined to be approximately constant at 0.035 within a frequency range of 0.3–2.2 THz, in comparison with the THz spectroscopy of a permalloy/Pt multilayer. Our results not only demonstrate the potential of L1$_2$-Mn$_3$Ir as a spintronic THz emitter but also provide insights into the THz spin transport properties of L1$_2$-Mn$_3$Ir.

A spintronic terahertz (THz) emitter is a device that emits single-cycle THz electromagnetic waves through ultrafast spin-to-charge conversion in a heavy metal layer coupled to a ferromagnetic metal.[1-3] The irradiation of a femtosecond laser pulse to the heterostructure triggers ultrafast demagnetization and flow of hot electrons inside the ferromagnetic layer, producing a spin current pulse with a sub-picosecond duration.[4,5] The spin current pulse is then converted into charge current, resulting in single-cycle THz electromagnetic wave emissions.[1-3] Spintronic THz emitters with Pt and W[2,6] are already commercialized and comparable to THz crystals in terms of the bandwidth and flexibility.[3] Because the efficiency of the THz-emission process depends on the spin Hall (SH) angle of the spin-to-charge conversion layer,[7] much effort is being made to develop spintronic THz emitters from new materials with larger SH angles and new functionalities.

Antiferromagnets have recently emerged as promising candidates for spintronic devices operating in the THz frequency range. Noncollinear antiferromagnets of Mn$_3$X (X = Sn, Ge, Pt, Ga, Ir, and Rh) exhibit large anomalous Hall and magneto-optical Kerr effects,[8-20] due to the spin-orbit coupling (SOC) and noncollinear spin textures.[19,20] Among these compounds, Mn$_3$Ir has a particularly high SH conductivity owing to its strong SOC.[21,22] Unlike the γ-phase disordered alloy $Mn_{100-x}Ir_x$,[23] the $L1_2$-ordered phase of $Mn_3Ir$ has a face-centered cubic lattice with an all-in/all-out triangular spin structure in the (111) lattice plane (Fig.1(a)) as a result of the competition between the magnetic frustration and exchange interactions.[24,25] The $L1_2$-ordered phase of $Mn_3Ir$ exhibits a high Néel temperature of ~960 K.[25] The ordered triangular magnetic configuration with such high thermal stability can play a major role in imposing an exchange bias on a ferromagnetic layer in a spin valve structure employed as a read head of a hard disk drive.[26] However, despite its potential, the capability of $L1_2$-ordered antiferromagnetic $Mn_3Ir$ as a spintronic THz emitter is yet to be explored, in contrast to conventional heterostructures that utilize nonmagnetic heavy metals.[27-30] In this paper, we report the observation of THz-wave emissions resulting from the ultrafast spin-to-charge conversion in an $L1_2$-$Mn_3Ir$ film. By comparing the THz emission spectra with the spectra of Pt, we quantified the SH angle of $L1_2$-$Mn_3Ir$ in the THz spectral range, which was determined to be almost constant at 0.035 up to a frequency of 2.2 THz.

We deposited a 15 nm-thick $L1_2$-ordered $Mn_3Ir$ film epitaxially grown on a MgO(111) substrate at 600 °C by direct current sputtering. Previous studies have demonstrated that $Mn_3Ir$ films prepared under similar deposition conditions exhibited a sizable anomalous Hall effect. [12,13,14] Moreover, we deposited a 3 nm-thick permalloy (Py) layer as the spin current source. The multilayer was protected from oxidation by a 5 nm-thick $SiO_2$ capping layer. We also prepared Py(3 nm)/Pt(5 nm) and Pt(3 nm)/$Mn_3Ir$(15 nm) on MgO(111) substrates as control samples using the same sputtering system. The crystal structures of the films were analyzed by X-ray diffraction (XRD). Figure 1(b) shows the out-of-plane $\theta$-$2\theta$ XRD profile of the Py/$Mn_3Ir$ multilayer, measured using $K\alpha_1$(Cu) X-ray source. The out-of-plane XRD profile without any secondary peaks indicates the epitaxial growth of the $Mn_3Ir$ film on the MgO (111) substrate. To evaluate the magnitude of $L1_2$ ordering of the $Mn_3Ir$ film, we measured the X-ray reflection of the $L1_2$-$Mn_3Ir$ (001) and (002) planes with a tilt angle of 54.7° with respect to the (111) plane. Notably, the X-ray reflection of the $L1_2$-$Mn_3Ir$ (001) and (002) planes correspond to the superlattice and fundamental diffraction peaks, respectively.[31] The results are shown in Fig. 1(c). The order parameter, $C$, was calculated using the following equation: [31,32]

$$C = \sqrt{\frac{I_{001}(f_{Ir} + 3f_{Mn})^2 LP(\theta_{002})A(\theta_{002})}{I_{002}(f_{Ir} - f_{Mn})^2 LP(\theta_{001})A(\theta_{001})}}, \qquad (1)$$

where $I_{001}$ and $I_{002}$ are the integrated values of the (001) and (002) peaks at a diffraction angle of $2\theta_{001}$ and $2\theta_{002}$, respectively. The structure factors of the superlattice peak, $f_{Ir} + 3f_{Mn}$, and the fundamental peak, $f_{Ir} - f_{Mn}$, were calculated using the atomistic scattering factors, $f_{Ir} = 77$ and $f_{Mn} = 25$, respectively. We also considered the angular dependences of the Lorentz-polarization factor, $LP(\theta) = (1 + \cos^2 2\theta)/\sin^2\theta\cos\theta$, and the absorption factor, $A(\theta) = (1 - e^{-\frac{2t\mu}{\sin\theta}})/2\mu$, with the absorption coefficient, $\mu = 0.251$ µm$^{-1}$. The order parameter calculated using Eq. (1) was $C = 0.33$. The calculated interplanar spacing for the (111) plane was $d_{111} = 0.2182$ nm, which was close to that of bulk Mn$_3$Ir, $d_{111} = 0.2181$ nm. By contrast, the lattice parameter along the oblique [001] direction was 0.3807 nm, which was larger than the bulk lattice parameter of 0.3778 nm. This indicates the presence of in-plane tensile strain by the deposition to the MgO (111) substrate due to the lattice mismatch of ~9 % between Mn$_3$Ir and MgO. We have also observed the twinning of the Mn$_3$Ir crystal in the XRD $\varphi$-scan data (Fig. 1(e)) in comparison with that of the MgO (002) peak (Fig. 1(d)). The twinning percentage determined from the $\varphi$ XRD scan of the (002) peak (Fig. 1(d)) was 34%.

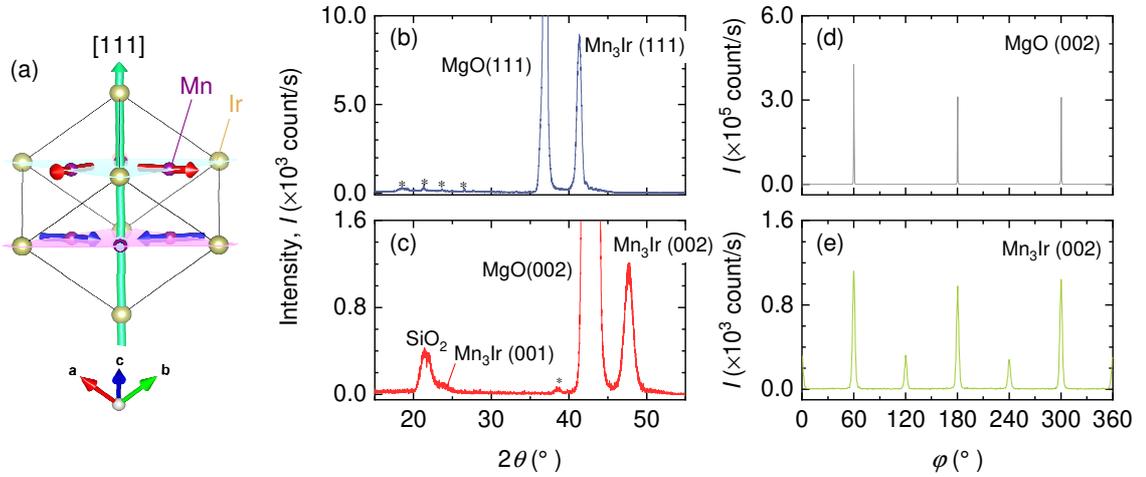

FIG. 1. (a) Crystal and spin structure of L1$_2$-Mn$_3$Ir. (b) $\theta$-$2\theta$ XRD pattern of the Py/Mn$_3$Ir/MgO (111) sub. (c) $\theta$-$2\theta$ X-ray reflection pattern of the same sample measured under a tilt angle of 54.7° from the direction normal to the (111) plane. XRD $\varphi$-scans of the (d) (002) diffraction peak of the MgO substrate and (e) (002) diffraction peak of the Mn$_3$Ir layer. Here, the asterisk symbols indicate the diffractions from a sample mount made of clay.

For the THz emission and transmission experiments, we employed a Yb: KGW laser system with a central wavelength, repetition frequency, and pulse width of 1028 nm, 10 kHz, and 230 fs, respectively. Using a permanent magnet, we applied an in-plane magnetic field of ±0.5 kOe to a multilayer for saturating the magnetization of the Py layer. The pump pulses were modulated using an optical chopper at 570 Hz, enabling the detection of pump-induced THz signals by a lock-in amplifier. We detected the THz waves through changes in the ellipticity of a probe pulse by the electro-optic effect of an 800 µm thick CdTe(110) crystal. The THz emission and detection processes were performed in a dry nitrogen environment at room temperature. See Ref. 30 for further details on the measurement configuration.

By exciting the sample with linearly polarized pulses, the spin currents flowed from the Py layer to the Mn$_3$Ir layer, generating THz waves (Fig. 2(a)). Figure 2(b) shows the THz emission signals, $S_{\text{emit}}(t)$, acquired when applying a magnetic field, $H = +0.5$ kOe and $-0.5$ kOe, to the Py/Mn$_3$Ir and Pt/Mn$_3$Ir multilayers. The polarity of the THz waves originating from the Py/Mn$_3$Ir multilayer was inverted when the $H$-direction was reversed. By contrast, no signal was observed for the Pt/Mn$_3$Ir multilayer. These results indicate that THz waves from the Py/Mn$_3$Ir multilayer are induced by spin currents resulting from ultrafast demagnetization of the Py layer. Contrastingly, the Mn$_3$Ir layer acts as an ultrafast spin-to-charge converter but not as a spin-current source, as similarly reported in to the case of Mn$_3$Sn[33]. In addition, the fluence dependence of the normalized peak intensity shows that the THz emission intensity, $S_{emit}(t)$, monotonically increases within the limit of the fluence range used in this study (Fig. 2(c)). To measure the time-reversal odd component of the SH angle,[17,21,34] we applied an out-of-plane magnetic field of ±140 kOe, which was sufficiently large to presumably obtain a minor hysteresis response associated with the anomalous Hall effect[13] in a similar Mn$_3$Ir film. After applying the magnetic field of ±140 kOe by using a superconducting magnet, we then measured $S_{emit}(t)$ using our THz emission set-up. However, we did not find any meaningful changes in $S_{emit}(t)$ before and after applying the magnetic field of ±140 kOe. This independence can be attributed to the small remanence magnetization, multidomain state, and crystal twinning of the Mn$_3$Ir film.

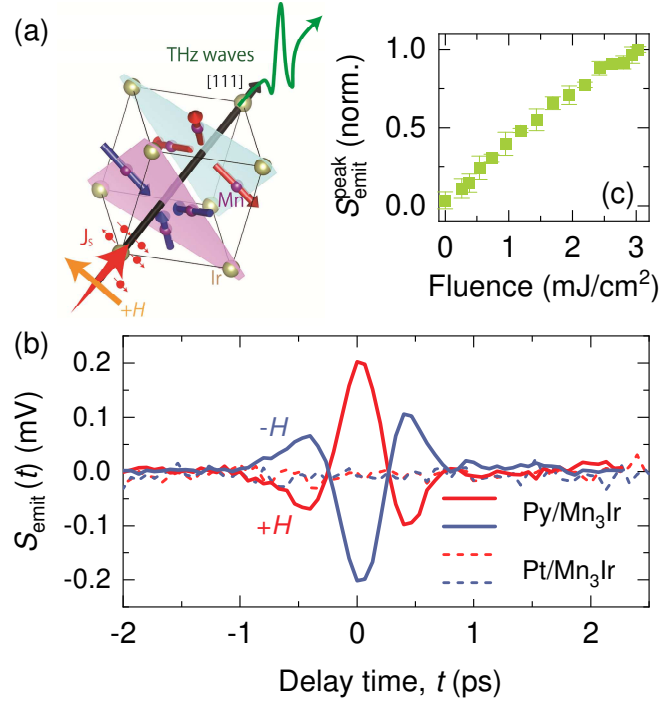

FIG. 2. (a) Schematic of the THz emission from spin-to-charge conversion inside the L1$_2$-Mn$_3$Ir. (b) THz-emission signals ($S_{emit}(t)$) of the Py/Mn$_3$Ir (solid lines) and Pt/Mn$_3$Ir (dashed lines) under an external magnetic field ($H$) of ±0.5 kOe. (b) Here, the pump fluence was set at 2.94 mJ/cm$^2$. (c) Pump fluence dependence of the peak intensity of $S_{emit}(t)$. The dependency was normalized based on the data obtained with a pump fluence of 2.94 mJ/cm$^2$.

The THz electric field, $\boldsymbol{E}(\omega)$, generated by the spin-to-charge conversion effects can be described in terms of the angular frequency domain as follows:[7,35,36]

$$E(\omega) \propto \tilde{T}(\omega) \frac{2e}{\hbar} \theta_{\text{SH}}(\omega) j_s^{\text{demag}}(\omega) \frac{\lambda_{\text{sd}}}{t_{\text{SC}}} \tanh \frac{t_{\text{SC}}}{2\lambda_{\text{sd}}} \mathbf{n} \times \boldsymbol{\sigma}, \qquad (2)$$

where $e$, $\hbar$, $\mathbf{n}$, and $\boldsymbol{\sigma}$ represent the electron charge, Dirac constant, unit vector normal to the film plane, and spin unit vector, respectively. The parameters $\theta_{\text{SH}}$, $\lambda_{\text{sd}}$, and $t_{\text{SC}}$ denote the SH angle, spin diffusion length, and thickness of the spin-to-charge conversion layer, respectively. The direction of the $\boldsymbol{\sigma}$ vector is parallel to the localized spin direction of the ferromagnetic layer. Here, we neglected the longitudinal spin-to-charge conversions. The density of spin current created through the demagnetization of the ferromagnetic layer, $j_s^{\text{demag}}(\omega)$, was assumed to be linearly proportional to the magnetization component parallel to an external magnetic field and the product of absorbance and fluence of excitation pulse, namely, $A \cdot P_{\text{pump}}$. The magnetizations of the Py/Mn$_3$Ir and Py/Pt multilayers are 531 ± 14 emu/cm$^3$ and 770 ± 11 emu/cm$^3$ at 0.5 kOe, respectively. The absorbance of the Py/Mn$_3$Ir and Py/Pt multilayers are $A$ = 0.512 and 0.408, respectively, which were estimated by measuring the transmission and reflectance of the samples. See Supplementary Material for details of estimating $A$.

The complex transmittance in the angular frequency domain, $\tilde{T}(\omega)$, was calculated by the following equation:[7,30,37]

$$\tilde{T}(\omega) = \frac{S_{\text{trans}}^{\text{film}}(\omega)}{S_{\text{trans}}^{\text{sub}}(\omega)} e^{-i\Delta\Phi}, \qquad (3)$$

where $S_{\text{trans}}(\omega)$ is the amplitude of the complex fast Fourier transform (FFT) of the THz transmission signal. The phase and amplitude changes according to the difference in substrate thickness, $\Delta d_{\text{sub}}$, are included in the phase, $\Delta\Phi = i(\tilde{n}_{\text{sub}}(\omega) - n_0)\Delta d_{\text{sub}}\omega/c$, where $\tilde{n}_{\text{sub}}(\omega)$, $n_0$, and $c$ denote the refractive index of the substrate, refractive index of air, and speed of light, respectively. Combining Eqs. (2) and (3), we can estimate the value of $\theta_{\text{SH}}$ by considering the results of the transmission experiments obtained for the control samples, Py/Pt multilayer, and bare MgO substrate. Figures 3(a) and (b) show the $S_{\text{trans}}(t)$ and FFT spectra, $|S_{\text{trans}}(\omega)|$, respectively. To estimate the $\Delta\Phi$ in Eq. (3), we used $n_0 = 1$, and the frequency dependence of $\tilde{n}_s(\omega)$ (Ref. 29). The actual measurement values of $\Delta d_{\text{sub}}$ were −0.031 mm and 0.007 mm for the Py/Mn$_3$Ir and Py/Pt multilayers, respectively.

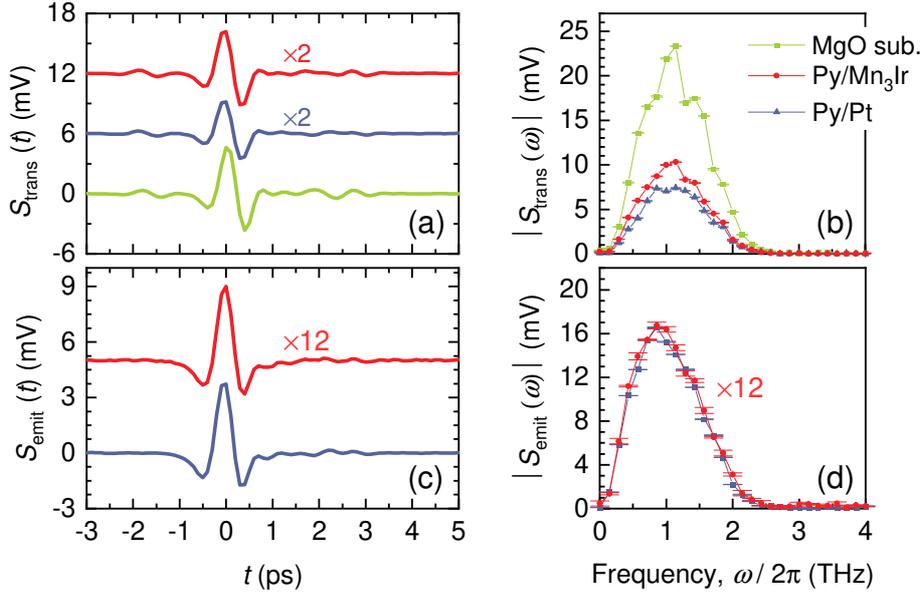

FIG. 3 (a) THz transmission signals ($S_{\text{trans}}(t)$) of the Py/Mn$_3$Ir multilayer (red), Py/Pt reference multilayer (blue), and bare MgO (111) substrate (light green). (b) Fast Fourier transform (FFT) spectra ($|S_{\text{trans}}(\omega)|$) of the samples. (c) THz emission signals and (d) FFT spectra ($|S_{\text{emit}}(\omega)|$) of the Py/Mn$_3$Ir and Py/Pt multilayers. The inset numbers indicate the multiplication factors. The error bars in the spectra were estimated from the standard errors of the signals.

The THz wave emission signals of the Py/Mn$_3$Ir and Py/Pt multilayers are shown in Fig. 3(c). To estimate the $\theta_{\text{SH}}$ values of the Mn$_3$Ir layer, the ratio of the FFT spectra in Fig. 3(d), $|S_{\text{emit}}(\omega)|$, and the real part of $\tilde{T}(\omega)$ were substituted in Eq. (2). The sign of $\theta_{\text{SH}}$ at each frequency was determined by taking into account of phase information obtained from the FFT. Here, we assumed that the following parameters were constant within the frequency range: $\theta_{\text{SH}} = 0.12$ and $\lambda_{\text{sd}} = 1.4$ nm for the Pt layer[38], and $\lambda_{\text{sd}} = 1.0$ nm for the Mn$_3$Ir layer.[39] We ignored the spin-to-charge conversion in the Py layer because of the small SH angle, i.e., $\theta_{\text{SH}} = 0.005$.[40] The $\theta_{\text{SH}}$ values in the THz frequency range of this work can then be analyzed by THz emission and transmission spectroscopy. The results of these analyses are shown in Fig. 4. The $\theta_{\text{SH}}$ values first increased within a frequency range of 0.15–0.30 THz, whereas they remained nearly constant at 0.035 up to a frequency of 2.2 THz. Ab initio calculations may give insight as to the reduction of spin Hall angle in the frequency dependence around 0.15 THz. We propose that the dispersion of $\theta_{\text{SH}}$ is effective to observe the modulation of the spin Hall effect by magnons[41,42] and phonons[43] of other antiferromagnets in the THz frequency range. In addition, the estimated spin Hall angle for our Mn$_3$Ir film, $\theta_{\text{SH}} = 0.035$, was smaller compared with those (0.10–0.15) obtained from transport experiments of (111)-oriented and polycrystalline disordered Mn$_3$Ir films.[39] Theoretical calculations predicted a large negative SH conductivity of L1$_2$-Mn$_3$Ir.[22,39] The net $\theta_{\text{SH}}$ of our Mn$_3$Ir film may be small because of the mixture of L1$_2$-ordered and disordered Mn$_3$Ir crystals with the spin Hall effects of opposite signs. Investigating the order-parameter dependence of the THz emission will be valuable to uncover the origin of the THz emission, which would lead us to further enhance the spin-conversion efficiency of Mn$_3$Ir systems.

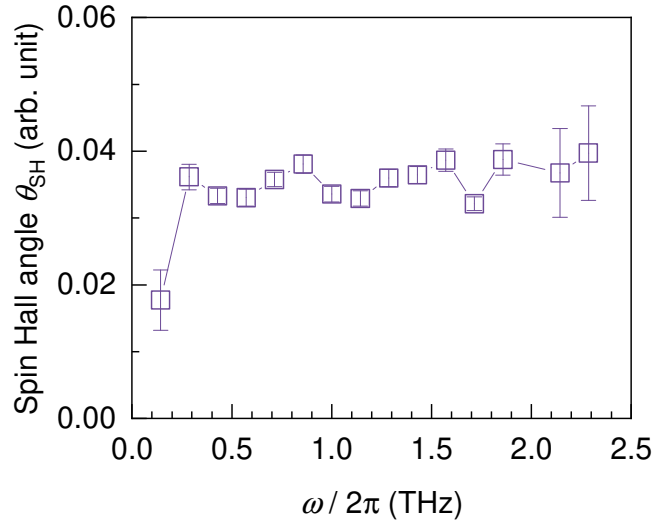

FIG. 4. Frequency dependence of spin Hall angle for the (111)-ordered $L1_2$-$Mn_3Ir$ film. The error bars were estimated from the standard errors of the THz transmission (Fig. 2(b)) and emission (Fig. 2(d)) spectra.

In this study, we investigated the THz emissions resulting from ultrafast spin-to-charge conversions via the inverse spin Hall effect in a (111)-oriented $L1_2$-$Mn_3Ir$ thin film on a MgO(111) substrate. Based on the XRD profiles, we found that the $Mn_3Ir$ layer had an $L1_2$ ordering of 0.33 and was distorted in the in-plane direction due to the lattice mismatch with the MgO(111) substrate. Our control experiments revealed that the $Mn_3Ir$ layer was not a spin source but a spin-to-charge converter in the current experimental configuration. By comparing the THz wave emission and transmission results for the Py/$Mn_3Ir$ multilayer with those for the Py/Pt multilayer, the spin Hall angle of the $Mn_3Ir$ layer was calculated to be ~0.035 within the THz frequency range investigated in this study. We believe that our results and methodology will be useful in the search for promising antiferromagnets for THz spintronic applications.

See the Supplementary Material for additional details on the ultrafast terahertz measurement setup, transmission and reflectance of the pump light, and analysis of the THz emission spectrum.

**AUTHOR DECLARATIONS**

**Conflict of interest**

The authors have no conflicts to disclose.

**DATA AVAILABILITY**

The data that support the findings of this study are available from the corresponding author upon reasonable request.

ACKNOWLEDGMENTS

We thank Prof. H. Munekata, Prof. T. Satoh, and Dr. D. Bossini for critically reading the manuscript and Dr. Y. Takamura and Prof. S. Nakagawa for technical guidance in the XRD measurements. This work was partly supported by JSPS KAKENHI (Grant nos. 22K14588, 21K14218, and 21H04562), Sasakawa Scientific Research Grant (Grant no. 2023-2032), JST PRESTO (Grant nos. JPMJCR22C3 and JPMJPR20B9), and the Collaborative Research Program of the Institute for Chemical Research, Kyoto University.

# Supplementary Material

Huiling Mao,[1] Yuta Sasaki,[2] Yuta Kobayashi,[3] Shinji Isogami,[2] Teruo Ono,[3,4] Takahiro Moriyama,[3,4] Yukiko K. Takahashi,[2] and Kihiro T. Yamada[1]

1) *Department of Physics, Tokyo Institute of Technology, Tokyo 152-8551, Japan*
2) *National Institute for Materials Science, Tsukuba, Ibaraki 987-6543, Japan*
3) *Institute for Chemical Research, Kyoto University, Uji, Kyoto 611-0011, Japan*
4) *Center for Spintronics Research Network, Institute for Chemical Research, Kyoto University, Uji, Kyoto, 611-0011, Japan*

A) **Measurement of absorption, reflection, and transmission**

Figures S1(a) and (b) show the absorptance, transmittance, and reflectance of the pump light as a function of pump power for the Py/Mn$_3$Ir and reference Py/Pt multilayers, respectively. The absorptance of the pump light was calculated by subtracting the power of the reflected and transmitted pump light from the total power.

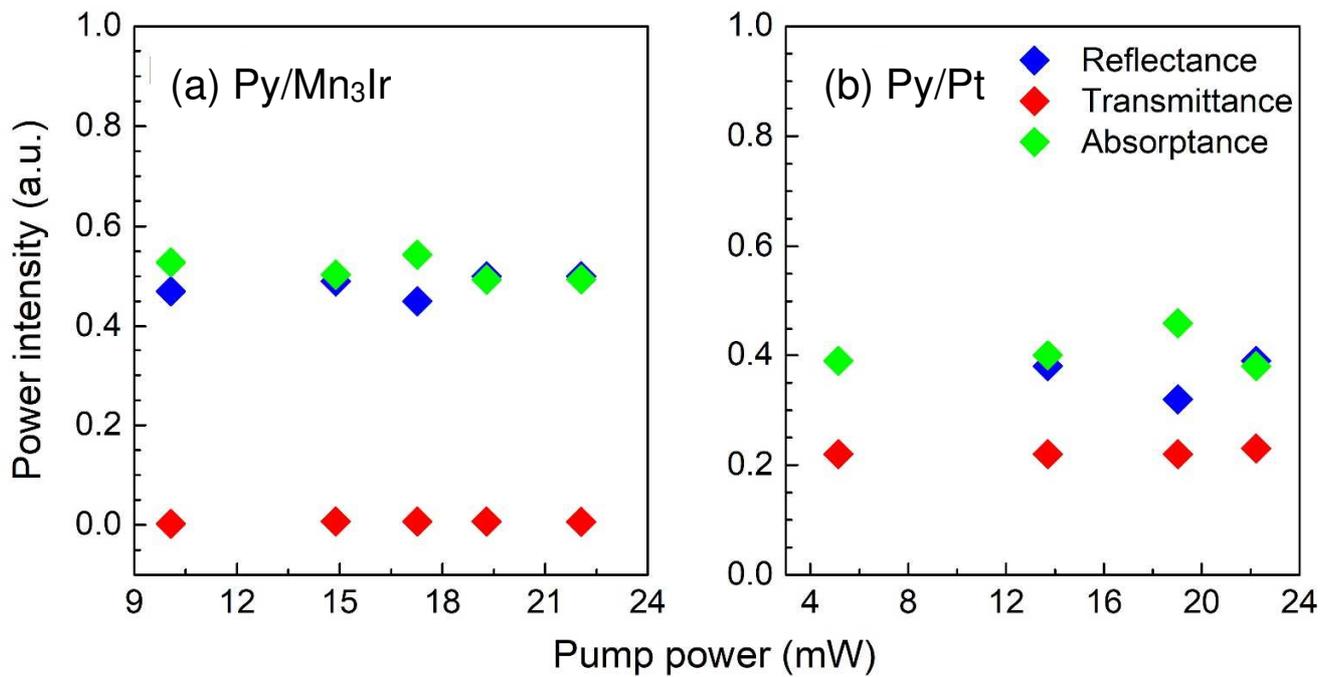

FIG. S1. Comparison of the pump light absorptance, transmittance, and reflectance of the Py/Mn$_3$Ir and Py/Pt multilayers. The power intensities were divided by the total pump power.